\documentclass[ 
aps,%
10pt,%
final,%
notitlepage,%
oneside,%
onecolumn,%
nobibnotes,%
nofootinbib,%
superscriptaddress,%
noshowpacs,%
centertags]%
{revtex4-2}

\usepackage[cp1251]{inputenc}
\usepackage[T2A]{fontenc}
\usepackage[dvips]{color}
\usepackage{amsmath,amsfonts,amssymb,amsbsy,epsf,colordvi,rotate,array}
\usepackage[mathscr]{eucal}
\usepackage{hyperref}
\usepackage{graphicx}
\begin{document}
	
\title{Cut-off scale of quantum gravity as a sign to unify interactions}
	
\begin{abstract}
Dynamics of gravity interaction with matter at one-loop level of effective quantum field theory naturally sets the cut-off scale $\Lambda_E$ in a sub-Planckian region through  incorporating the gauge coupling constant $\alpha(\Lambda_E)$ and the reduced Planckian mass $\tilde m_\textsc{pl}$ into relation $ \Lambda_E\sim \tilde m_\textsc{pl}\,\alpha(\Lambda_E)/(4\pi)$ numerically yielding $\Lambda_E\sim 10^{16}$ GeV. Such two scales are empirically inherent for an inflationary cosmology.
\end{abstract}

	\author{\firstname{Asya}~\surname{Aynbund}}
	\email{aynbund.asya@phystech.edu}
	\affiliation{Moscow Institute of Physics and Technology,
		Institutsky 9, Dolgoprudny, Moscow Region, 141701, Russia}
	
	\author{\firstname{V.V.}~\surname{Kiselev}}
	\email{kiselev.vv@phystech.edu; Valery.Kiselev@ihep.ru}
	
	\affiliation{Moscow Institute of Physics and Technology,
		Institutsky 9, Dolgoprudny, Moscow Region, 141701, Russia}
	\affiliation{Institute for High Energy Physics named by A.A.~Logunov of National
Research Centre «Kurchatov Institute», Nauki 1, Protvino, Moscow Region, 142281, Russia}	
	\maketitle
\section{Introduction}	
General relativity perfectly describes the non-linear interaction of geometry with matter in a region wherein  quantum fluctuations in the gravity sector are negligible. To exhibit a limit of such a classical approximation by the order of magnitude one usually considers the scalar curvature $R$ as setting the inverse length $\lambda_\textsc{gr}$ squared 
\begin{equation}
\label{naiv-1}
	R\sim \frac{1}{\lambda_\textsc{gr}^2},
\end{equation}
so that $\lambda_\textsc{gr}$ is treated to be a quantum wave length of gravitational field. Einstein equations in terms of Ricci tensor $R_{\mu\nu}$ of metrics $g_{\mu\nu}$
\begin{equation}
\label{Ei}
	R_{\mu\nu}-\frac12\,g_{\mu\nu}\,R=8\pi G\,T_{\mu\nu}
\end{equation}
relate the geometry with the stress-energy tensor $T_{\mu\nu}$ of matter, wherein $8\pi G=1/\tilde m_\textsc{pl}^{2}$ expresses the Newton constant $G$ through the reduced Planckian mass $ \tilde m_\textsc{pl}\approx 2.4\cdot 10^{18}$ GeV. 
The components of matter stress-energy tensor is evaluated in terms of energy scale $\Lambda$  setting  $T_{00}\sim \Lambda^4$. Therefore, by the order of magnitude one finds 
\begin{equation}
\label{naiv-2}
	R\sim \frac{1}{\lambda_\textsc{gr}^2} \sim \frac{\Lambda^4}{\tilde m_\textsc{pl}^{2}}\sim 8\pi G\,T\quad 
	\Rightarrow\quad 
	 \frac{1}{\lambda_\textsc{gr}}\sim \Lambda\,\frac{\Lambda}{\tilde m_\textsc{pl}}.
\end{equation}
Quantum fluctuation of energy  $\delta E_\textsc{gr}$ in the gravity sector  obeys the uncertainty relation (in units $\hbar=c=1$)
\begin{equation}
\label{ur}
	\delta E_\textsc{gr}\cdot \lambda_\textsc{gr }\sim 1,
\end{equation}
hence, 
\begin{equation}
\label{naiv-3}
	\delta E_\textsc{gr}\sim \Lambda\,\frac{\Lambda~}{\tilde m_\textsc{pl}}.
\end{equation}
In this way, quantum gravity effects are negligible if fluctuations $\delta E_\textsc{gr}$ are much less than the energy scale $\Lambda$ set by the matter
$$
	\delta E_\textsc{gr}\ll \Lambda,
$$
while the quantum gravity cut-off is reached if $\delta E_\textsc{gr}\sim \Lambda$ that gives 
\begin{equation}
\label{naiv-4}
	\Lambda\sim \tilde m_\textsc{pl}.
\end{equation}
Analogous estimates are discussed in fresh researches reviewed in \cite{Castellano:2024bna}, while other ideas are presented in \cite{tHooft:1993dmi,Hossenfelder:2012jw,Rovelli:2004tv,Rovelli:1997yv}. 

However, the initial suggestion of (\ref{naiv-1}) relating the scalar curvature $R$ with the quantum gravitational length 
$\lambda_\textsc{gr }$ crucially changes if one involves arguments coming from cutting off integrals at one-loop level. 

\begin{figure}
\begin{center}
\includegraphics[width=2.5in]{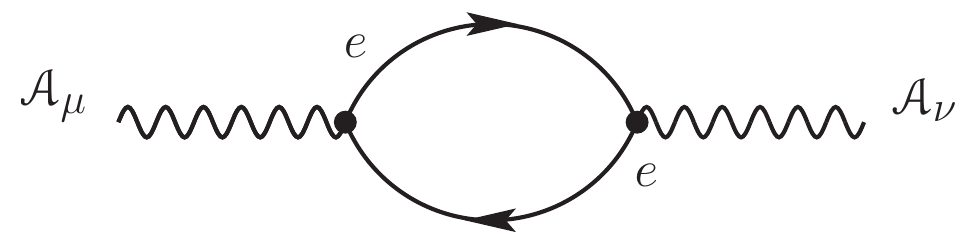}
\caption{One-loop contribution to the kinetic term of external source $\mathscr A_\mu$ coupled to the current $\bar\psi\,\gamma^\mu\psi$ with the charge $e$.}
\label{QED-1}
\end{center}
\end{figure}

For the sake of clarity in speculations let us start with a free Dirac spinor field $\psi$ possessing the  action 
\begin{equation}
\label{D-psi}
	S_\textsc{d}=\int\mathrm d^4 x \,\bar \psi\,(-\mathrm i \gamma^\mu\partial_\mu-m)\,\psi.
\end{equation}
Add the external source $\mathscr A_\mu$ in the standard interaction with the conserved current $j^\mu=\bar\psi\,\gamma^\mu\psi$ and charge $e$
\begin{equation}
\label{D-int}
	S_\mathrm{int}=-e\int \mathrm d^4 x \,\big(\bar \psi\,\gamma^\mu\psi\big)\cdot\mathscr A_\mu.
\end{equation}
The one-loop quantum correction to the effective action as shown in FIG. \ref{QED-1} generates the kinetic term of source $\mathscr A_\mu$. 
Due to the current conservation the integration in Euclidian momenta after the Wick rotation symbolically results in the  term alike 
\begin{equation}
\label{QED-2}
	\frac12\,\mathscr A_\mu \mathscr A_\nu \big(-g^{\mu\nu}k^2+k^\mu k^\nu\big)\cdot e^2
	\int \frac{\mathrm d^4 p_E}{(2\pi)^4}\,\frac{1}{p_E^4}.
\end{equation}
Since the norm of integration 
\begin{equation}
\label{int-E}
	\int \frac{\mathrm d^4 p_E}{(2\pi)^4}=
	\int\frac{\mathrm d\Omega_3}{2(2\pi)^4} \int {p_E^2\mathrm d p^2_E}=
	\frac{2\pi^2}{2(2\pi)^4}\int {p_E^2\mathrm d p^2_E}=\frac{1}{(4\pi)^2}\int {p_E^2\mathrm d p^2_E},
\end{equation}
one gets the kinetic term with the factor of the order of 
$$
	\left(\frac{e}{4\pi}\right)^2.
$$
Due to the renormalisability of quantum electrodynamics one can introduce the kinetic counter-term with the appropriate normalisation, of course. Nevertheless, we conclude that the involvement of sources for conserved currents allows us to analyse the scales in quantum loops. 

The similar procedure is described in Section \ref{sec-2} in the case of conserved stress-energy tensor to find a relevant cut-off in a non-renormalisable effective theory of matter interacting with metrics. The analysis in Section \ref{sec-2} is used in Section \ref{sec-3} to modify relations (\ref{naiv-1})--(\ref{naiv-4}). In Section \ref{sec-4} we argue for origin of cut-off in the quantum gravity with charge unified with charges of all interactions. Results are summarised in Conclusion. 

\section{Loop for kinetic term of metric\label{sec-2}}
\begin{figure}
\begin{center}
\includegraphics[width=2.5in]{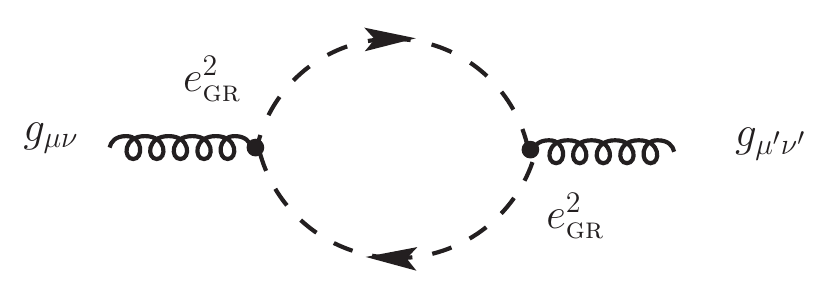}
\caption{One-loop contribution to the kinetic term of external source related with metric $g_{\mu\nu}$ and coupled to the current $T^{\mu\nu}$ with the charge $e^2_\textsc{gr}$.}
\label{GR-1}
\end{center}
\end{figure}

Consider a free real scalar field $\phi$ possessing conserved stress-energy tensor
\begin{equation}
\label{scal-1}
	T_{\mu\nu}=\partial_\mu\phi\,\partial_\nu\phi-g_{\mu\nu}\frac12\big((\partial_\lambda\phi)^2-m^2\phi^2\big).
\end{equation}
This current interacts with external source of inverse metric $g^{\mu\nu}$.  The interval written down in tangent Minkowski space-time
$$
	\mathrm d s^2=\eta_{ab}\mathrm d y^a\mathrm d y^b,\qquad a,b\in\{\overline{0,3}\},
	\qquad \eta_{ab}\equiv\mbox{daig}(+1,-1,-1,-1),
$$
can be transformed to the word coordinates $x^\mu$ (see details on classical and quantum gravity in status review \cite{Carlip:2001wq})
$$
	\mathrm d s^2=\eta_{ab}\frac{\partial y^a}{\partial x^\mu}\,\frac{\partial y^b}{\partial x^\nu}\,
	\mathrm d x^\mu \mathrm d x^\nu=g_{\mu\nu}\mathrm d x^\mu \mathrm d x^\nu,
$$
so that the metric is the composite field of two co-tetrads  $\mathfrak e^a_\mu={\partial y^a}/{\partial x^\mu}$
\begin{equation}
\label{tetrad}
	g_{\mu\nu}=\eta_{ab}\, \mathfrak e^a_\mu \mathfrak e^b_\nu.
\end{equation}
In general relativity the tetrad (vierbein) 
$$
	\mathfrak e_a^\mu=\frac{\partial x^\mu}{\partial y^a}$$ is related with the co-tetrad 
$$
	\mathfrak e_a^\mu \mathfrak e^b_\mu=\delta_a^b,\qquad 
	\mathfrak e_a^\mu \mathfrak e^a_\nu =\delta^\mu_\nu,
$$
so that the co-tetrad is torsion free 
$$
	\partial_\nu\mathfrak e^a_\mu-\partial_\mu\mathfrak e^a_\nu\equiv 0,
$$
and the connection determining the covariant derivative is consistent with the metric. So, we postulate that the co-tetrad $\mathfrak e^a_\mu$ corresponds to the primordial field $\epsilon^a_\mu$  in gravity so that the canonical dimension of  the field equals 
$[\epsilon^a_\mu]=[E]$. In general relativity the stress-energy tensor appears in the interaction with the metric to the leading order 
\begin{equation}
\label{tetrad-2b}
	\delta S_\mathrm{int}=\frac12   \int\mathrm d^4 x\,\delta g_{\mu\nu}\, T^{\mu\nu}=\frac12
	 \int\mathrm d^4 x\, \delta(\eta_{ab}\, \mathfrak e^a_\mu \mathfrak e^b_\nu)\, T^{\mu\nu}.
\end{equation}
The interaction of dynamical field $\epsilon^a_\mu$ with the conserved current $T_a^\mu$ involves the charge $e_\textsc{gr}$ in the Lagrangian 
$$
	\mathscr L_\mathrm{int}\sim e_\textsc{gr}\,T_a^\mu\,\epsilon^a_\mu.
$$
However, the true canonical dimension of Lagrangian requires the introduction of factor $1/\Lambda_E$ with $[\Lambda_E]=[E]$, where $\Lambda_E$ denotes the only underlying scale separating the quantum gravity from the effective theory, that is the cut-off. So, combining the definition of stress-energy tensor in (\ref{tetrad-2b}) with  dimensional transition from the co-tetrad to the dynamical field
\begin{equation}
\label{dyn-tetrad}
	 \mathfrak e^a_\mu\;\mapsto\;e_\textsc{hr}\,\epsilon^a_\mu\,\frac{1}{\Lambda_E}
\end{equation}
we get 
 the interaction of stress-energy tensor with external source 
\begin{equation}
\label{tetrad-2}
	S_\mathrm{int}=\frac12\,  e_\textsc{gr} \int\mathrm d^4 x\,T_a^\mu\,\epsilon^a_\mu\,\frac{1}{\Lambda_E}=
	\frac12\,  e_\textsc{gr} \int\mathrm d^4 x\,T^{\mu\nu}\,
	\mathfrak e^b_\nu\,\eta_{ab}\,\epsilon^a_\mu\,\frac{1}{\Lambda_E}=
	\frac12\,  e_\textsc{gr}^2 \int\mathrm d^4 x\,T^{\mu\nu}\,
	\epsilon^b_\nu\,\eta_{ab}\,\epsilon^a_\mu\,\frac{1}{\Lambda_E^2}.
\end{equation}
Denoting the composite metric field
$$
	\tilde h_{\mu\nu}\equiv \epsilon^b_\nu\,\eta_{ab}\,\epsilon^a_\mu
$$
with dimension $[\tilde h_{\mu\nu}]=[E]^2$ by curled curve we show the one-loop diagram contributing to the kinetic term of such the metric in FIG. \ref{GR-1}.
According to (\ref{int-E}) the kinetic term of metric acquires the charge factor of 
\begin{equation}
\label{gr-2}
	\left(\frac{e_\textsc{gr}^2}{4\pi}\right)^2,
\end{equation}
while the momentum integral diverges quadratically in cut-off $\Lambda_E$
$$
	\int\frac{\mathrm d p_E^2}{p_E^4}\,p_E^4 \sim \Lambda_E^2.
$$
Thus, the kinetic term
\begin{equation}
\label{kin-gr}
	S_\textsc{gr}^\mathrm{kin}\sim \frac12\int\mathrm d^4 x\,
	 \big(\tilde h_{\mu\nu}\,k^2\, \tilde h_{\mu'\nu'}\big)\,\Pi^{\mu\nu\mu'\nu'}\,
	 \frac{1}{\Lambda_E^2} \left(\frac{e_\textsc{gr}^2}{4\pi}\right)^2
\end{equation}
includes the massless spin-2 matrix of transverse polarisations\footnote{Standard transverse projector $P^{\mu\nu}\equiv g^{\mu\nu}-k^\mu k^\nu/k^2$ determines $\Pi^{\mu\nu\mu'\nu'}=P^{\mu\mu'}P^{\nu\nu'}+P^{\mu\nu'}P^{\nu\mu'}-P^{\mu\nu}P^{\mu'\nu'}.$} $\Pi$, while the dimensional factor has to be reduced to an appropriate form in the leading order of the Einstein--Hilbert action in general relativity. So, all of traces by the coupling constant as well as the cut-off should disappear to get expansions in the inverse Planckian mass $\tilde m_\textsc{pl}$. Therefore, 
\begin{equation}
\label{planck}
	\frac{1}{\Lambda_E^2} \left(\frac{e_\textsc{gr}^2}{4\pi}\right)^2\sim \frac{1}{\tilde m_\textsc{pl}^2}
\end{equation}
or for the canonical spin-2 field $h_{\mu\nu}=\tilde h_{\mu\nu}/\tilde m_\textsc{pl}$ we get 
\begin{equation}
\label{h}
	S_\textsc{gr}^\mathrm{kin}\sim \frac12\int\mathrm d^4 x\,
	 \big( h_{\mu\nu}\,k^2\,  h_{\mu'\nu'}\big)\,\Pi^{\mu\nu\mu'\nu'}.
\end{equation}
We suppose that a grand unification dictates 
\begin{equation}
\label{GU}
	e_\textsc{gr}^2\equiv \alpha_\textsc{gu}(\Lambda_E)
\end{equation}
in Gauss units. Therefore, we have substanciated that
\begin{equation}
\label{cut-main}
	\Lambda_E\sim \tilde m_\textsc{pl}\cdot \frac{\alpha_\textsc{gu}(\Lambda_E)}{4\pi}.
\end{equation}

\section{Fluctuations of gravitational energy\label{sec-3}}
Relation (\ref{cut-main}) implies that the connection of scalar curvature $R$ to the gravitational wave length  $\lambda_\textsc{gr}$ is modified to
\begin{equation}
\label{r-lambda}
	R\sim \left( \frac{\alpha_\textsc{gu}(\Lambda)}{4\pi}\,\frac{1}{\lambda_\textsc{gr}}\right)^2\sim 
	\frac{\Lambda^4}{\tilde m_\textsc{pl}^{2}}.
\end{equation}
Therefore, the uncertainty relation (\ref{ur}) gives the estimate of energy fluctuations
\begin{equation}
\label{deltaE}
	\delta E_\textsc{gr}\sim \Lambda\,\frac{\Lambda~}{\tilde m_\textsc{pl}}\cdot 
	\frac{4\pi}{\alpha_\textsc{gu}(\Lambda)}.
\end{equation}
Numerical estimate of quantum gravity cut-off at $\delta E_\textsc{gr}\sim \Lambda$ yields
\begin{equation}
\label{num}
	\Lambda_E\sim \tilde m_\textsc{pl}\,\frac{\alpha_\textsc{gu}(\Lambda_E)}{4\pi},\qquad
	\Lambda_E\sim 10^{16}\mbox{ GeV,}\qquad \alpha_\textsc{gu}(\Lambda_E)\sim \frac{1}{25}.
\end{equation}
At the energy density restricted by $\Lambda\ll \Lambda_E$ any quantum gravity effects are negligible.

\section{Origin of cut-off\label{sec-4}}
Loop corrections generate stress-energy densities and fields at scales governed by the maximal 4D virtualities. These virtualities curve the space-time. A scale of curvature as the inverse radius $r_E$ depends on both the virtuality $\Lambda_E$ and charge $e_\textsc{gr}$ of gravitational interaction with matter fields. In accordance with dimensional arguments denote
\begin{equation}
\label{rE-1}
	\frac{1}{r_E}=\tilde \alpha\,\Lambda_E,
\end{equation}
so that we expect that the coefficient $\tilde \alpha$ diminishes if the charge tends to zero,
\begin{equation}
\label{rE-2}
	\tilde \alpha\to 0\quad\mbox{if}\quad e_\textsc{gr}\to 0.
\end{equation}
To the leading order the energy density $\Lambda_E^4$ inside radius $r_E$ generates mass $M_E$,
\begin{equation}
\label{rE-3}
	M_E=\frac{4\pi}{3}\,\Lambda_E^4 r_E^3,
\end{equation}
that creates the Schwarzschild black hole of radius $r_g$,
$$
	r_g=2 G M_E= \frac{1}{4\pi \tilde m_\textsc{pl}^2}\,\frac{4\pi}{3}\,\Lambda_E^4 r_E^3\sim 
	 \frac{\Lambda_E^2}{\tilde m_\textsc{pl}^2}\,\big(\Lambda_E r_E\big)^2r_E.
$$
Therefore, one could neglect the black hole formation if $r_g<r_E$, while at the scale of cut-off in quantum gravity one reaches
\begin{equation}
\label{rE-4}
	r_g\sim r_E\quad\Rightarrow\quad 
	 \frac{1}{\Lambda_E r_E}\sim\frac{\Lambda_E}{\tilde m_\textsc{pl}}.
\end{equation}
Hence, the scale of curvature is related to the ratio of cut-off to the Planckian mass,
\begin{equation}
\label{rE-5}
	\frac{1}{r_E}=\tilde \alpha\Lambda_E\sim \frac{\Lambda_E}{\tilde m_\textsc{pl}}\,\Lambda_E.
\end{equation}
At $e_\textsc{gr}\to 0$ one has got
$$
	\frac{\Lambda_E}{\tilde m_\textsc{pl}}\to 0.
$$
These speculations do not involve any quantum-loops estimates, hence, one cannot establish a form of $\tilde \alpha$ dependence on $e_\textsc{gr}$, of course. However, in previous sections we have explored one-loop arguments to set up
\begin{equation}
\label{rE-6}
	\tilde\alpha\sim\frac{e_\textsc{gr}^2}{4\pi}. 
\end{equation}
Thus, the cut-off scale of quantum gravity involving the charge has to be sub-Planckian, while the cut-off itself appears due to forming black holes which are excluding any wave lengths below a horizon radius in accordance with (\ref{rE-4}). 

To enforce the arguments consider the black hole of mass $M_\textsc{bh}$ possessing horizon radius $r_g$ in general relativity. The black hole is formed due to accretion of matter to the surface $A_\textsc{bh}=4\pi r_g^2$. At the forming of black hole the mass occupies a narrow layer of thickness $r_E$ in vicinity of the horizon surface. So, the density of energy in such a layer equals
\begin{equation}
\label{rE-7}
	\rho_M=\frac{M_\textsc{bh}}{4\pi r_g^2 r_E}.
\end{equation}
Substituting the gravitational radius
$$
	r_g\sim \frac{M_\textsc{bh}}{4\pi \tilde m_\textsc{pl}^2},
$$
one arrives to 
\begin{equation}
\label{rE-8}
	\rho_M \sim 4\pi \frac{\tilde m_\textsc{pl}^4}{M_\textsc{bh} r_E}.
\end{equation}
If the density of mass $\rho_M$ does not exceed a critical value $\Lambda_E^4$ determined by the cut-off scale in the quantum gravity with charge $e_\textsc{gr}$, then the black hole admits a pure classical description in the framework of general relativity at 
\begin{equation}
\label{rE-9}
	M_\textsc{bh} \gg \frac{\tilde m_\textsc{pl}^4}{\Lambda_E^4 r_E}\sim 
	 \frac{\tilde m_\textsc{pl}}{\tilde \alpha^2}.
\end{equation}
The critical mass is reached at $r_g\sim r_E$ again yielding (\ref{rE-4}). This makes the confidence to the substantiation of cut-off value at sub-Planckian scale and its connection to the charge unification of all known kinds of interactions in nature. 

\section{Conclusion}
We have derived the dynamical connection of two separate scales inherent for the gravity: the Planckian mass and the cut-off in quantum loops (see also relevant indications concerning for the scale of cut-off in \cite{Balitsky:2014epa,Aynbund:2024lby,Aynbund:2025jza}). Analogous two scales first have appeared empirically in the theory of inflationary expansion of early Universe as unrelated parameters \cite{Starobinsky:1980te,Guth:1980zm,Linde:1981mu,Linde:1983gd,Albrecht:1982mp,Bardeen:1983qw}: the Planckian mass and the energy scale in a flat plateau of inflaton potential \cite{Kallosh:2013tua,Galante:2014ifa,Kallosh:2025ijd}. 
Remark that the plateau of inflaton potential is placed in the very exit from the quantum gravity to the effective theory. 
Henceforth these scales have acquired the common origin while the gravity has joined the grand unification with all of interactions as exhibited by the fusion of coupling constants into a single value at the scale of gravity cut-off. 

We expect that analogous speculations for a conserved super-current in the supersymmetric field theory would result in a correct form of kinetic term for massless gravitino. 

In our derivation we have not traced for the sign in front of kinetic term since it depends on complete set of bosonic and fermionic fields in a model.\\[-5mm]

\begin{acknowledgments}
The work of Asya Aynbund was supported by grant <<PhD Student>> of <<BASIS>> Foundation for Development of Theoretical Physics and Mathematics.
\end{acknowledgments}

\bibliography{bib_Grav-cut}

\begin{thebibliography}{18}%
\makeatletter
\providecommand \@ifxundefined [1]{%
 \@ifx{#1\undefined}
}%
\providecommand \@ifnum [1]{%
 \ifnum #1\expandafter \@firstoftwo
 \else \expandafter \@secondoftwo
 \fi
}%
\providecommand \@ifx [1]{%
 \ifx #1\expandafter \@firstoftwo
 \else \expandafter \@secondoftwo
 \fi
}%
\providecommand \natexlab [1]{#1}%
\providecommand \enquote  [1]{``#1''}%
\providecommand \bibnamefont  [1]{#1}%
\providecommand \bibfnamefont [1]{#1}%
\providecommand \citenamefont [1]{#1}%
\providecommand \href@noop [0]{\@secondoftwo}%
\providecommand \href [0]{\begingroup \@sanitize@url \@href}%
\providecommand \@href[1]{\@@startlink{#1}\@@href}%
\providecommand \@@href[1]{\endgroup#1\@@endlink}%
\providecommand \@sanitize@url [0]{\catcode `\\12\catcode `\$12\catcode
  `\&12\catcode `\#12\catcode `\^12\catcode `\_12\catcode `\%12\relax}%
\providecommand \@@startlink[1]{}%
\providecommand \@@endlink[0]{}%
\providecommand \url  [0]{\begingroup\@sanitize@url \@url }%
\providecommand \@url [1]{\endgroup\@href {#1}{\urlprefix }}%
\providecommand \urlprefix  [0]{URL }%
\providecommand \Eprint [0]{\href }%
\providecommand \doibase [0]{https://doi.org/}%
\providecommand \selectlanguage [0]{\@gobble}%
\providecommand \bibinfo  [0]{\@secondoftwo}%
\providecommand \bibfield  [0]{\@secondoftwo}%
\providecommand \translation [1]{[#1]}%
\providecommand \BibitemOpen [0]{}%
\providecommand \bibitemStop [0]{}%
\providecommand \bibitemNoStop [0]{.\EOS\space}%
\providecommand \EOS [0]{\spacefactor3000\relax}%
\providecommand \BibitemShut  [1]{\csname bibitem#1\endcsname}%
\let\auto@bib@innerbib\@empty
\bibitem [{\citenamefont {Castellano}(2024)}]{Castellano:2024bna}%
  \BibitemOpen
  \bibfield  {author} {\bibinfo {author} {\bibfnamefont {A.}~\bibnamefont
  {Castellano}},\ }\emph {\bibinfo {title} {{The Quantum Gravity Scale and the
  Swampland}}},\ \href@noop {} {Ph.D. thesis},\ \bibinfo  {school} {U.
  Autonoma, Madrid (main)} (\bibinfo {year} {2024}),\ \Eprint
  {https://arxiv.org/abs/2409.10003} {arXiv:2409.10003 [hep-th]} \BibitemShut
  {NoStop}%
\bibitem [{\citenamefont {'t~Hooft}(1993)}]{tHooft:1993dmi}%
  \BibitemOpen
  \bibfield  {author} {\bibinfo {author} {\bibfnamefont {G.}~\bibnamefont
  {'t~Hooft}},\ }\bibfield  {title} {\bibinfo {title} {{Dimensional reduction
  in quantum gravity}},\ }\href@noop {} {\bibfield  {journal} {\bibinfo
  {journal} {Conf. Proc. C}\ }\textbf {\bibinfo {volume} {930308}},\ \bibinfo
  {pages} {284} (\bibinfo {year} {1993})},\ \Eprint
  {https://arxiv.org/abs/gr-qc/9310026} {arXiv:gr-qc/9310026} \BibitemShut
  {NoStop}%
\bibitem [{\citenamefont {Hossenfelder}(2013)}]{Hossenfelder:2012jw}%
  \BibitemOpen
  \bibfield  {author} {\bibinfo {author} {\bibfnamefont {S.}~\bibnamefont
  {Hossenfelder}},\ }\bibfield  {title} {\bibinfo {title} {{Minimal Length
  Scale Scenarios for Quantum Gravity}},\ }\href
  {https://doi.org/10.12942/lrr-2013-2} {\bibfield  {journal} {\bibinfo
  {journal} {Living Rev. Rel.}\ }\textbf {\bibinfo {volume} {16}},\ \bibinfo
  {pages} {2} (\bibinfo {year} {2013})},\ \Eprint
  {https://arxiv.org/abs/1203.6191} {arXiv:1203.6191 [gr-qc]} \BibitemShut
  {NoStop}%
\bibitem [{\citenamefont {Rovelli}(2004)}]{Rovelli:2004tv}%
  \BibitemOpen
  \bibfield  {author} {\bibinfo {author} {\bibfnamefont {C.}~\bibnamefont
  {Rovelli}},\ }\href {https://doi.org/10.1017/CBO9780511755804} {\emph
  {\bibinfo {title} {{Quantum gravity}}}},\ Cambridge Monographs on
  Mathematical Physics\ (\bibinfo  {publisher} {Univ. Pr.},\ \bibinfo {address}
  {Cambridge, UK},\ \bibinfo {year} {2004})\BibitemShut {NoStop}%
\bibitem [{\citenamefont {Rovelli}(1998)}]{Rovelli:1997yv}%
  \BibitemOpen
  \bibfield  {author} {\bibinfo {author} {\bibfnamefont {C.}~\bibnamefont
  {Rovelli}},\ }\bibfield  {title} {\bibinfo {title} {{Loop quantum gravity}},\
  }\href {https://doi.org/10.12942/lrr-1998-1} {\bibfield  {journal} {\bibinfo
  {journal} {Living Rev. Rel.}\ }\textbf {\bibinfo {volume} {1}},\ \bibinfo
  {pages} {1} (\bibinfo {year} {1998})},\ \Eprint
  {https://arxiv.org/abs/gr-qc/9710008} {arXiv:gr-qc/9710008} \BibitemShut
  {NoStop}%
\bibitem [{\citenamefont {Carlip}(2001)}]{Carlip:2001wq}%
  \BibitemOpen
  \bibfield  {author} {\bibinfo {author} {\bibfnamefont {S.}~\bibnamefont
  {Carlip}},\ }\bibfield  {title} {\bibinfo {title} {{Quantum gravity: A
  Progress report}},\ }\href {https://doi.org/10.1088/0034-4885/64/8/301}
  {\bibfield  {journal} {\bibinfo  {journal} {Rept. Prog. Phys.}\ }\textbf
  {\bibinfo {volume} {64}},\ \bibinfo {pages} {885} (\bibinfo {year} {2001})},\
  \Eprint {https://arxiv.org/abs/gr-qc/0108040} {arXiv:gr-qc/0108040}
  \BibitemShut {NoStop}%
\bibitem [{\citenamefont {Balitsky}\ and\ \citenamefont
  {Kiselev}(2014)}]{Balitsky:2014epa}%
  \BibitemOpen
  \bibfield  {author} {\bibinfo {author} {\bibfnamefont {J.~V.}\ \bibnamefont
  {Balitsky}}\ and\ \bibinfo {author} {\bibfnamefont {V.~V.}\ \bibnamefont
  {Kiselev}},\ }\bibfield  {title} {\bibinfo {title} {{Quantum origin of
  suppression for vacuum fluctuations of energy}},\ }\href
  {https://doi.org/10.1103/PhysRevD.90.125017} {\bibfield  {journal} {\bibinfo
  {journal} {Phys. Rev. D}\ }\textbf {\bibinfo {volume} {90}},\ \bibinfo
  {pages} {125017} (\bibinfo {year} {2014})},\ \Eprint
  {https://arxiv.org/abs/1406.3046} {arXiv:1406.3046 [gr-qc]} \BibitemShut
  {NoStop}%
\bibitem [{\citenamefont {Aynbund}\ and\ \citenamefont
  {Kiselev}(2024)}]{Aynbund:2024lby}%
  \BibitemOpen
  \bibfield  {author} {\bibinfo {author} {\bibfnamefont {A.}~\bibnamefont
  {Aynbund}}\ and\ \bibinfo {author} {\bibfnamefont {V.~V.}\ \bibnamefont
  {Kiselev}},\ }\bibfield  {title} {\bibinfo {title} {{Cosmological Constant
  Suppression in Non-Stationary Scalar Covariant State}},\ }\href@noop {} {\
  (\bibinfo {year} {2024})},\ \Eprint {https://arxiv.org/abs/2411.16181}
  {arXiv:2411.16181 [hep-th]} \BibitemShut {NoStop}%
\bibitem [{\citenamefont {Aynbund}\ and\ \citenamefont
  {Kiselev}(2025)}]{Aynbund:2025jza}%
  \BibitemOpen
  \bibfield  {author} {\bibinfo {author} {\bibfnamefont {A.}~\bibnamefont
  {Aynbund}}\ and\ \bibinfo {author} {\bibfnamefont {V.~V.}\ \bibnamefont
  {Kiselev}},\ }\bibfield  {title} {\bibinfo {title} {{Scalar Field Action
  under 4D Isotropic Cut-off and its Cosmological Impact}},\ }\href@noop {} {\
  (\bibinfo {year} {2025})},\ \Eprint {https://arxiv.org/abs/2501.05274}
  {arXiv:2501.05274 [hep-th]} \BibitemShut {NoStop}%
\bibitem [{\citenamefont {Starobinsky}(1980)}]{Starobinsky:1980te}%
  \BibitemOpen
  \bibfield  {author} {\bibinfo {author} {\bibfnamefont {A.~A.}\ \bibnamefont
  {Starobinsky}},\ }\bibfield  {title} {\bibinfo {title} {{A New Type of
  Isotropic Cosmological Models Without Singularity}},\ }\href
  {https://doi.org/10.1016/0370-2693(80)90670-X} {\bibfield  {journal}
  {\bibinfo  {journal} {Phys. Lett. B}\ }\textbf {\bibinfo {volume} {91}},\
  \bibinfo {pages} {99} (\bibinfo {year} {1980})}\BibitemShut {NoStop}%
\bibitem [{\citenamefont {Guth}(1981)}]{Guth:1980zm}%
  \BibitemOpen
  \bibfield  {author} {\bibinfo {author} {\bibfnamefont {A.~H.}\ \bibnamefont
  {Guth}},\ }\bibfield  {title} {\bibinfo {title} {{The Inflationary Universe:
  A Possible Solution to the Horizon and Flatness Problems}},\ }\href
  {https://doi.org/10.1103/PhysRevD.23.347} {\bibfield  {journal} {\bibinfo
  {journal} {Phys. Rev. D}\ }\textbf {\bibinfo {volume} {23}},\ \bibinfo
  {pages} {347} (\bibinfo {year} {1981})}\BibitemShut {NoStop}%
\bibitem [{\citenamefont {Linde}(1982)}]{Linde:1981mu}%
  \BibitemOpen
  \bibfield  {author} {\bibinfo {author} {\bibfnamefont {A.~D.}\ \bibnamefont
  {Linde}},\ }\bibfield  {title} {\bibinfo {title} {{A New Inflationary
  Universe Scenario: A Possible Solution of the Horizon, Flatness, Homogeneity,
  Isotropy and Primordial Monopole Problems}},\ }\href
  {https://doi.org/10.1016/0370-2693(82)91219-9} {\bibfield  {journal}
  {\bibinfo  {journal} {Phys. Lett. B}\ }\textbf {\bibinfo {volume} {108}},\
  \bibinfo {pages} {389} (\bibinfo {year} {1982})}\BibitemShut {NoStop}%
\bibitem [{\citenamefont {Linde}(1983)}]{Linde:1983gd}%
  \BibitemOpen
  \bibfield  {author} {\bibinfo {author} {\bibfnamefont {A.~D.}\ \bibnamefont
  {Linde}},\ }\bibfield  {title} {\bibinfo {title} {{Chaotic Inflation}},\
  }\href {https://doi.org/10.1016/0370-2693(83)90837-7} {\bibfield  {journal}
  {\bibinfo  {journal} {Phys. Lett. B}\ }\textbf {\bibinfo {volume} {129}},\
  \bibinfo {pages} {177} (\bibinfo {year} {1983})}\BibitemShut {NoStop}%
\bibitem [{\citenamefont {Albrecht}\ \emph {et~al.}(1982)\citenamefont
  {Albrecht}, \citenamefont {Steinhardt}, \citenamefont {Turner},\ and\
  \citenamefont {Wilczek}}]{Albrecht:1982mp}%
  \BibitemOpen
  \bibfield  {author} {\bibinfo {author} {\bibfnamefont {A.}~\bibnamefont
  {Albrecht}}, \bibinfo {author} {\bibfnamefont {P.~J.}\ \bibnamefont
  {Steinhardt}}, \bibinfo {author} {\bibfnamefont {M.~S.}\ \bibnamefont
  {Turner}},\ and\ \bibinfo {author} {\bibfnamefont {F.}~\bibnamefont
  {Wilczek}},\ }\bibfield  {title} {\bibinfo {title} {{Reheating an
  Inflationary Universe}},\ }\href
  {https://doi.org/10.1103/PhysRevLett.48.1437} {\bibfield  {journal} {\bibinfo
   {journal} {Phys. Rev. Lett.}\ }\textbf {\bibinfo {volume} {48}},\ \bibinfo
  {pages} {1437} (\bibinfo {year} {1982})}\BibitemShut {NoStop}%
\bibitem [{\citenamefont {Bardeen}\ \emph {et~al.}(1983)\citenamefont
  {Bardeen}, \citenamefont {Steinhardt},\ and\ \citenamefont
  {Turner}}]{Bardeen:1983qw}%
  \BibitemOpen
  \bibfield  {author} {\bibinfo {author} {\bibfnamefont {J.~M.}\ \bibnamefont
  {Bardeen}}, \bibinfo {author} {\bibfnamefont {P.~J.}\ \bibnamefont
  {Steinhardt}},\ and\ \bibinfo {author} {\bibfnamefont {M.~S.}\ \bibnamefont
  {Turner}},\ }\bibfield  {title} {\bibinfo {title} {{Spontaneous Creation of
  Almost Scale - Free Density Perturbations in an Inflationary Universe}},\
  }\href {https://doi.org/10.1103/PhysRevD.28.679} {\bibfield  {journal}
  {\bibinfo  {journal} {Phys. Rev. D}\ }\textbf {\bibinfo {volume} {28}},\
  \bibinfo {pages} {679} (\bibinfo {year} {1983})}\BibitemShut {NoStop}%
\bibitem [{\citenamefont {Kallosh}\ \emph {et~al.}(2014)\citenamefont
  {Kallosh}, \citenamefont {Linde},\ and\ \citenamefont
  {Roest}}]{Kallosh:2013tua}%
  \BibitemOpen
  \bibfield  {author} {\bibinfo {author} {\bibfnamefont {R.}~\bibnamefont
  {Kallosh}}, \bibinfo {author} {\bibfnamefont {A.}~\bibnamefont {Linde}},\
  and\ \bibinfo {author} {\bibfnamefont {D.}~\bibnamefont {Roest}},\ }\bibfield
   {title} {\bibinfo {title} {{Universal Attractor for Inflation at Strong
  Coupling}},\ }\href {https://doi.org/10.1103/PhysRevLett.112.011303}
  {\bibfield  {journal} {\bibinfo  {journal} {Phys. Rev. Lett.}\ }\textbf
  {\bibinfo {volume} {112}},\ \bibinfo {pages} {011303} (\bibinfo {year}
  {2014})},\ \Eprint {https://arxiv.org/abs/1310.3950} {arXiv:1310.3950
  [hep-th]} \BibitemShut {NoStop}%
\bibitem [{\citenamefont {Galante}\ \emph {et~al.}(2015)\citenamefont
  {Galante}, \citenamefont {Kallosh}, \citenamefont {Linde},\ and\
  \citenamefont {Roest}}]{Galante:2014ifa}%
  \BibitemOpen
  \bibfield  {author} {\bibinfo {author} {\bibfnamefont {M.}~\bibnamefont
  {Galante}}, \bibinfo {author} {\bibfnamefont {R.}~\bibnamefont {Kallosh}},
  \bibinfo {author} {\bibfnamefont {A.}~\bibnamefont {Linde}},\ and\ \bibinfo
  {author} {\bibfnamefont {D.}~\bibnamefont {Roest}},\ }\bibfield  {title}
  {\bibinfo {title} {{Unity of Cosmological Inflation Attractors}},\ }\href
  {https://doi.org/10.1103/PhysRevLett.114.141302} {\bibfield  {journal}
  {\bibinfo  {journal} {Phys. Rev. Lett.}\ }\textbf {\bibinfo {volume} {114}},\
  \bibinfo {pages} {141302} (\bibinfo {year} {2015})},\ \Eprint
  {https://arxiv.org/abs/1412.3797} {arXiv:1412.3797 [hep-th]} \BibitemShut
  {NoStop}%
\bibitem [{\citenamefont {Kallosh}\ and\ \citenamefont
  {Linde}(2025)}]{Kallosh:2025ijd}%
  \BibitemOpen
  \bibfield  {author} {\bibinfo {author} {\bibfnamefont {R.}~\bibnamefont
  {Kallosh}}\ and\ \bibinfo {author} {\bibfnamefont {A.}~\bibnamefont
  {Linde}},\ }\bibfield  {title} {\bibinfo {title} {{On the Present Status of
  Inflationary Cosmology}},\ }\href@noop {} {\  (\bibinfo {year} {2025})},\
  \Eprint {https://arxiv.org/abs/2505.13646} {arXiv:2505.13646 [hep-th]}
  \BibitemShut {NoStop}%
\end{thebibliography}%

\end{document}